\documentstyle[prb,aps,12pt]{revtex}
\begin{document}
\draft
\title{Instanton in Disordered Peierls Systems}
\author{Maxim V. Mostovoy\cite{Perm} and Jasper Knoester}
\address{Institute for Theoretical Physics, Materials Science Center\\
University of Groningen, Nijenborgh 4, 9747 AG Groningen, The Netherlands}
\date{\today}
\maketitle

\begin{abstract}

We study disordered Peierls systems described by the Fluctuating
Gap Model.  We show that the typical electron states with
energies lying deep inside the pseudogap are localized near
large disorder fluctuations (instantons), which have the form of
a soliton-antisoliton pair.  Using the ``saddle-point'' method we
obtain the average density of states and the average optical
absorption coefficient at small energy.

\end{abstract}


\hspace{1.1cm}keywords: Peierls systems, disorder, instanton

\vspace{1cm}

Disorder in quasi-one-dimensional conductors strongly affects the
electron states, reducing the tendency of these materials to
develop $2 k_F$-instabilities, such as the Peierls
instability.\cite{Peierls} Existing Peierls materials, like the
conjugated polymer trans-polyacetylene, are known to suffer from
various kinds of disorder: conformational defects, cross-links,
impurities, {\em etc}.\cite{Kies} While in a perfect Peierls
chain the single-electron spectrum has a gap, with a value
related to the amplitude of the periodic chain distortion, in a
disordered chain the gap is filled.  At weak disorder, the
density of disorder-induced states that occur inside the gap is
small, leading to a pseudogap, while at large disorder the gap
disappears entirely and in the middle of the band the density may
even diverge.\cite{OE}

In this Letter, we find the typical form of the disorder-induced
electron states lying close to the center of the pseudogap.  In
addition to giving new insight into these disordered systems,
this result is useful, as it allows for a relatively easy
calculation of disorder averages.  As examples, we will
explicitly calculate the average density of states and the
average absorption coefficient at small energies.

It is crucial for our approach, that a large disorder fluctuation
is required to create an electron state close to the center of
the pseudogap, implying that the probability for such a state to
occur is small at weak disorder.  The main contribution to the
density of states at small energy then comes from the disorder
realizations close to one most probable fluctuation.  The form of
this fluctuation can be found consistently with the form of the
wave function of the electron state induced inside the pseudogap.
This wave function turns out to be localized in the vicinity of
the disorder fluctuation.  A similar situation is encountered
when calculating the density of states of electrons moving in a
random potential at large negative energies.\cite{HL,ZL} These
states were found to be localized in regions where the disorder
potential has the form of a deep well.

The approximate calculation of disorder averages, valid when the
dominant contribution comes from a small part of all possible
disorder realizations, is closely related to the semiclassical
approximation in quantum mechanics and field theory.  In the path
integral version of this approximation, the paths giving the
largest contribution to the Green function lie close to the
``saddle-point'' of the Euclidean action, called the
instanton.\cite{Coleman}

To describe the disordered Peierls chain we use the Fluctuating
Gap Model (FGM).\cite{Keldysh} It has previously been considered
in the context of the thermodynamical properties of
quasi-one-dimensional organic compounds (NMP-TCNQ,
TTF-TCNQ),\cite{OE} and has been applied to study the effect of
disorder on the Peierls transition,\cite{XT} as well as the
effect of quantum lattice fluctuations on the optical spectrum of
Peierls materials.\cite{Kim,H&M,S&M} In this model the electron
motion is described by the one-dimensional Dirac equation,
\begin{equation}
\label{Dirac}
{\hat h} \psi = \left( \sigma_3 \frac{v_F}{i} \frac{d}{dx}
+ \sigma_1 \Delta(x) \right) \psi(x) = \varepsilon \psi(x) \;,
\end{equation}
where $\sigma_1$ and $\sigma_3$ are the Pauli matrices and
\[ \psi(x) = \left( \begin{array}{c}
\psi_{R}(x) \\ \psi_{L}(x) \\ \end{array} \right) \]
is the wave function of the single-electron state close to the
Fermi energy $\varepsilon_F = 0$.  The first term in the
Hamiltonian ${\hat h}$ describes the free motion of the electrons
and the two amplitudes $\psi_{R}(x)$ and $\psi_{L}(x)$ correspond
to particles moving, respectively, to the right and to the left
with the Fermi velocity $v_F$.  The second term in the
Hamiltonian describes the backward scattering of electrons from
the lattice distortion wave, whose amplitude is proportional to
$\Delta(x)$.  Disorder is modelled by assuming $\Delta(x)$ to
fluctuate randomly along the chain around some average value
$\Delta_0$,
\begin{equation}
\label{Delta}
\Delta(x) = \Delta_0 + \eta(x) \;,
\end{equation}
where $\eta(x)$ is the fluctuating part with a
Gaussian correlator,
\begin{equation}
\label{Gauss}
\langle \eta (x) \eta(y) \rangle = A \delta(x - y) \;.
\end{equation}

In the absence of disorder ($\Delta(x) = \Delta_0$), the electron
spectrum has a gap between the energies $\varepsilon=-\Delta_0$
and $\varepsilon=+\Delta_0$.  At nonzero disorder the average
density of states lying close to the middle of the pseudogap
($|\varepsilon| \ll \Delta_{0}$) was found in Ref.\CITE{OE} by
means of the ``phase formalism'',\cite{LSP}
\begin{equation}
\label{dos}
\langle
\rho(\varepsilon) \rangle \propto
|\varepsilon|^{\frac{2}{g} - 1} \;,
\end{equation}
where $g = A \left/ \left( v_F
\Delta_0 \right) \right.$.  We note that due to the charge
conjugation symmetry (particle-hole symmetry) of the Dirac
Hamiltonian ${\hat h}$, the density of states is a symmetric
function of the energy
$
\langle \rho(- \varepsilon) \rangle =
\langle \rho( \varepsilon) \rangle
$,
and in what follows we will assume $\varepsilon$ to be
positive.

From Eq.(\ref{dos}) it is clear that at weak disorder ($g \ll 1$)
the density of states close to the middle of the pseudogap is
strongly suppressed.  As we mentioned above, the explanation for
this is that a large fluctuation of $\Delta(x)$ is required in
order to create an electron state with energy $\varepsilon \ll
\Delta_{0}$.  This motivates us to apply the ``saddle-point''
approach to study the typical electron states and to calculate
disorder averages.  The ``saddle-point''disorder fluctuation (or
instanton) $\bar{\eta}(x)$ is the least suppressed one among the
required large fluctuations.  It can be found by minimizing,
\begin{equation}
\label{Action} {\cal A}\left[ \eta(x) \right] =
\frac{1}{2A} \int\!\!  dx \eta^{2}(x) - {\mu} \left(
\varepsilon_{+}\left[ \eta(x) \right] - \varepsilon \right) \;.
\end{equation}
The first term in this equation describes the suppression of the
probability of the fluctuation with the correlator
Eq.~(\ref{Gauss}) (the weight $p[\eta(x)]$ of the disorder
configuration is $\exp \left( -\frac{1}{2A} \int\!\!dx\eta^{2}(x)
\right)$), while the second term stems from the condition that
the energy $\varepsilon_{+}\left[\eta(x) \right]$ of the lowest
positive energy single-electron state for the disorder realization
$\eta(x)$ equals $\varepsilon$.  The factor $\mu$ is a Lagrange
multiplier.  The minimization of ${\cal A}\left[ \eta(x) \right]$
gives,
\begin{equation}
\label{Saddle}
\bar{\eta}(x) = \mu A
\psi_+^{\dag}(x) \sigma_{1} \psi_{+}(x) \;,
\end{equation}
where $\psi_+(x)$ is the wave function of the state with energy
$\varepsilon_{+}\left[ \bar{\eta}(x) \right]$.

It may be shown by inspection that the solution of
Eq.(\ref{Saddle}) is a soliton-antisoliton pair configuration,
\begin{equation}
\label{Soan}
\bar{\eta}(x) = - v_F K
\left[
\tanh\left( K(x - x_0 + \frac{R}{2}) \right) -
\tanh\left( K(x - x_0 - \frac{R}{2}) \right)
\right] ,
\end{equation}
where $x_0$ describes the position of the disorder fluctuation
in the chain, $R$ is the distance between the soliton and the
antisoliton, and $K$ is determined by
\begin{equation}
\label{VFK}
v_F K = \Delta_0 \tanh (KR)  \;.
\end{equation}
The instanton is shown in Fig.~1 by plotting $\Delta(x) =
\Delta_0 + \bar{\eta}(x)$.  The spectrum of electron states that
occur for this $\Delta(x)$ has previously been
considered\cite{BK,CB} in relation with polarons in the
Su-Schrieffer-Heeger model of conjugated polymers,\cite{SSH,TLM}
and is depicted in Fig.~2.  It consists of a valence band (with
highest energy $- \Delta_{0}$), a conduction band (with lowest
energy $\Delta_{0}$), and two localized intragap states with
energies $\pm \varepsilon_+(R)$, where
\begin{equation}
\label{epsilon}
\label{EPS}
\varepsilon_+(R) =  \frac{\Delta_{0}}{\cosh (KR)} \;.
\end{equation}
Thus, the soliton-antisoliton separation $R$ is fixed by the
condition $\varepsilon_+(R) = \varepsilon$.  The two intragap
states are the bonding and antibonding superpositions of the
midgap states localized near the soliton and antisoliton:
\begin{equation}
\label{psi}
\psi_{\pm R}(x) = \psi_{\pm L}^{\ast}(x) = \sqrt{\frac{K}{8}}
\left[
\frac{e^{i \frac{\pi}{4}}}
{\cosh \left( K(x - x_0 - \frac{R}{2}) \right)}
\pm \frac{e^{- i \frac{\pi}{4}}}
{\cosh \left( K(x - x_0 + \frac{R}{2}) \right)}
\right] \;,
\end{equation}
($\psi_{\pm}^{\dagger}(x) \psi_{\pm}(x)$ is schematically plotted
in Fig.1).  The energy splitting $2 \varepsilon$ decreases
exponentially with the soliton-antisoliton separation, so that
for $\varepsilon \ll \Delta_0$,
\begin{equation}
\label{dist}
R \approx  \xi_0 \ln \frac{2 \Delta_0}{\varepsilon} \;,
\end{equation}
where $\xi_0 = v_F / \Delta_0$ is the correlation length.
The suppression factor Eq.(\ref{Action}),
\begin{equation}
{\cal A}\left[ \bar{\eta}(x) \right] \approx
\frac{1}{2A} (2 \Delta_0)^2 R \approx \frac{2}{g}
\ln \frac{2 \Delta_0}{\varepsilon}\;\;,
\end{equation}
also depends logarithmically on energy, so that the weight of the
saddle-point configuration is,
\begin{equation}
\label{weight}
p\left[ \bar{\eta}(x) \right] \propto
\varepsilon^{\frac{2}{g}}\;.
\end{equation}
This result already gives a good estimate for the shape of the
density of states inside the pseudogap at $g \ll 1$ (cf.
Eq.(\ref{dos})).

A more detailed calculation requires performing the Gaussian
integration over the disorder
realizations close to the ``saddle-point'' configuration
Eq.(\ref{Soan}).
We then find the
following expression for the average density of electron states
per unit length,
\begin{equation}
\label{ansdos}
\langle \rho (\varepsilon) \rangle =
\frac{e}{\pi g v_F}
\left( \frac{e \varepsilon}{2 \Delta_0}
\right)^{\frac{2}{g} - 1}\;.
\end{equation}
For $g \ll 1$, this agrees with Eq.(36) of Ref.\CITE{OE},
confirming the validity of the ``saddle-point'' approximation at
small energies and weak disorder.  The easiest way to get the
result Eq.(\ref{ansdos}) is to use the correspondence between the
averaging over disorder realizations $\eta(x)$ and the
quantum-mechanical averaging over the ground state for a certain
double-well potential.  The details of the calculation will be
reported elsewhere.

Having obtained the form of the most probable disorder-induced
electron states, we can now also calculate in a relatively
straightforward way the optical absorption coefficient for a
half-filled chain at photon energy $\omega \ll 2 \Delta_0$ and $g
\ll 1$.  Again, only a large disorder fluctuation can make the
energy difference between the empty and filled electron levels
small.  With the highest probability the photon absorption will
induce a transition from the highest occupied to the lowest
unoccupied electron state.  Due to the particle-hole symmetry,
the energy of the lowest unoccupied state at half-filling should
equal $+\omega / 2$ , while the energy of the highest doubly
occupied state should equal $- \omega / 2$.  Hence, the
``saddle-point'' disorder configuration, whose probability
largely determines the absorption rate, is given by
Eqs.(\ref{Soan}), (\ref{VFK}), and (\ref{EPS}) with
$\varepsilon_+(R) = \omega/2$.

Thus, in the ``saddle-point'' approximation, the absorption
coefficient is the product of the averaged density of states
(which is essentially the probability to find the necessary
disorder fluctuation) and the strength of the optical transition
between the two intragap levels:
\begin{equation}
\label{abs}
\langle \alpha(\omega) \rangle = \langle C \omega
\sum_f \left| \langle f | \hat{d} | 0 \rangle \right|^2
\delta (E_f - E_0 - \omega) \rangle \approx
\frac{C \omega}{2}
\left| \langle + | {\hat d} | - \rangle \right|^2
\langle \rho (\varepsilon=\omega/2) \rangle \;.
\end{equation}
Here ${\hat d}$ is the electric dipole operator, $| \pm \rangle$
denote the wave functions of the intragap states (with energies
$\pm \omega / 2$) and $C$ is an $\omega$-independent coefficient
(for small $\omega$, we can neglect the weak $\omega$-dependence
of the real part of the dielectric constant).

The wave functions of the intragap states $| \pm \rangle$ are the
bonding and antibonding superpositions of the wave functions of
the midgap states localized near the soliton and the antisoliton
(see Eq.(\ref{psi})), from which the transition dipole matrix
element is obtained as,
\begin{equation}
\langle + | {\hat d} | - \rangle = \frac {q R}{2}\;.
\end{equation}
Here $q$ denotes the electron charge and $R$ is the
soliton-antisoliton separation.  Thus, for the asymptotic
behavior of the averaged absorption coefficient at low photon
energy we obtain,
\begin{equation}
\label{absorp}
\langle \alpha(\omega) \rangle \propto \omega^{\frac{2}{g}}
\left( \ln \frac{4 \Delta_0}{\omega} \right)^2\;.
\end{equation}

At this point we want to comment on the calculations of the
optical conductivity in Refs.  \CITE{Kim} and \CITE{S&M}, in
which the factorization approximation, $\langle G G \rangle =
\langle G \rangle \langle G \rangle$, was used to evaluate the
disorder average of the product of two Green functions.  From
above, it is clear that this approximation is not valid at low
photon energies, as it results in the optical conductivity (as
well as the absorption coefficient) being proportional to the
second power of the weight Eq.(\ref{weight}), rather than the
first (cf.  Eq.(\ref{abs})).  Of course, at weak disorder the
absorption at photon energies $\omega \ll 2 \Delta_0$ is small
anyhow, but the factorization approximation makes it even much
smaller.

We would also like to point out that the small difference
between the energies of the bonding and antibonding states is
potentially dangerous for the saddle-point calculation of both
the average optical absorption coefficient and density of states.
The problem arises in the calculation of the contribution of the
disorder realizations close to the ``saddle-point'' fluctuation:
\[
\eta(x) = {\bar \eta}(x) + \delta \eta(x)\;.
\]
The perturbation $\delta {\hat h} = \sigma_1 \delta \eta(x)$ can,
in principle, strongly mix the bonding and antibonding states,
because of the small energy denominator $2 \varepsilon$ appearing
in the perturbation series.  Such mixing would affect the values
of both the energy splitting between the two states and the
dipole matrix element.  To see if that is the case, we considered
the effective perturbation Hamiltonian, acting on the subspace of
the two bonding and antibonding states, which includes the
virtual excitations to all other (high-energy) electron states.
We found that the off-diagonal matrix elements of this
Hamiltonian are $O(\varepsilon)$, which cancels $\varepsilon$
in the denominator and makes the mixing of the two states small.
This result is a direct consequence of the charge conjugation
symmetry of the Dirac Hamiltonian Eq.(\ref{Dirac}).  Thus,
despite the small energy splitting, the saddle-point method is
applicable.

We conclude that in the FGM the most probable form of the wave
function of the electron state lying deep within the pseudogap
contains two peaks.  The ``saddle-point'' disorder fluctuation,
which induces such a state, has the form of a soliton-antisoliton
pair and the peaks of the wave function are localized near the
two kinks of this fluctuation (see Fig.~1).  Away from the kinks,
the electron wave function Eq.(\ref{psi}) falls off exponentially
on a length scale $\xi_0$.  This observation is consistent with
the fact, that the localization length at zero energy,
\[
l_{loc}(\varepsilon=0)=\frac{\xi_0}{(1-\frac{g}{2})}\;,
\]
calculated using the Thouless formula,\cite{Thouless} at weak
disorder equals the correlation length $\xi_0$.  As we
demonstrated, the instanton approach allows for a relatively easy
calculation of the small-energy density of states and absorption
coefficient.

Our results (Eqs.(\ref{ansdos}) and Eq.(\ref{absorp})) are valid
if the density of the disorder-induced states is small, which is
the case when $|\varepsilon| \ll \Delta_0$ and $g \ll 1$.  It is
useful, however, to comment briefly on effects of large disorder.
For $g \sim 1$ the typical size of the disorder fluctuation
$\eta(x)$ on a scale of the correlation length $\xi_0$ becomes
comparable to $\Delta_0$, so that disorder fluctuations inducing
the electron states with small energy are no longer suppressed,
and for $g > 2$ the density of states even diverges at
$\varepsilon = 0$.\cite{OE} This is essentially the singularity
found long ago by Dyson\cite{Dyson} for a gapless system, because
at strong disorder there is no principal difference between the
electron states in Peierls insulators and conducting ($\Delta_0 =
0$) chains.  In the latter case the localization length of the
electron states diverges as $\varepsilon \rightarrow 0$ and the
wave functions have an irregular structure, being large in many
separated chain regions.\cite{ER} Surprisingly, the
``saddle-point'' method gives the correct exponent $(2/g-1)$
(cf.~Eq.(\ref{ansdos})) for the energy dependence of the average
density of states at {\em all} values of $g$.  This suggests,
that even at large $g$ the typical form of the wave function may
be close to the one given by Eq.(\ref{psi}) in the regions where
the wave function is large, and that multi-instanton disorder
configurations (a gas of the soliton-antisoliton pairs) may
become important at strong disorder.

This work is supported by the "Stichting Scheikundig Onderzoek in
Nederland (SON)" and the "Stichting voor Fundamenteel Onderzoek
der Materie (FOM)".

\newpage

\section*{Figure Captions}

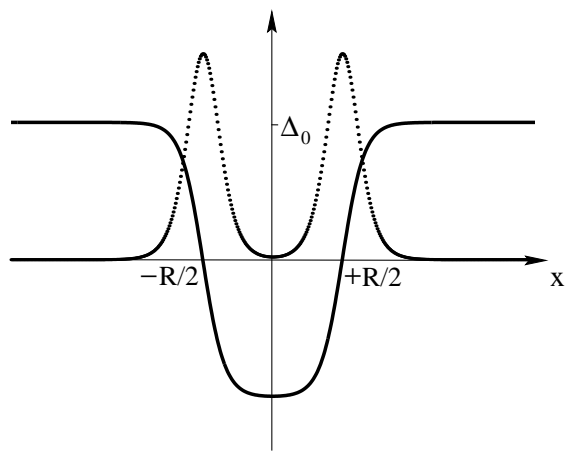
FIG.~1.  The form of $\Delta(x) = \Delta_0 + \bar{\eta}(x)$ for
the instanton disorder fluctuation (thick line) and the electron
density $|\psi_+(x)|^2 = |\psi_-(x)|^2$ for the corresponding
intragap states (dotted line).

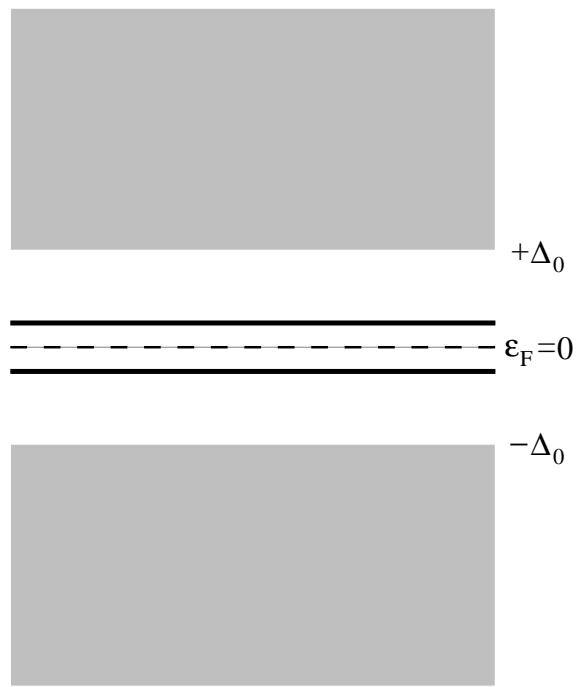
FIG.~2. The spectrum of electron states for the instanton
configuration $\Delta(x)$ plotted in Fig.~1.


\begin{references}
\bibitem[*]{Perm} Permanent address: Budker Institute of Nuclear
Physics, Novosibirsk, 630090, Russia.


\bibitem{Peierls} R.~E.~Peierls, {\em Quantum Theory of Solids},
(Clarendon, Oxford 1955), p.108.

\bibitem{Kies} see for a review, {\em Conjugated Conducting
Polymers}, ed.  H.~Kies, (Springer, Berlin, 1992).

\bibitem{OE}  A. A. Ovchinnikov and N. S. Erikhman,  Zh. Eksp.
Theor. Fiz. {\bf 73}, 650 (1977) [Sov. Phys. JETP {\bf 46}, 340
(1977)].

\bibitem{HL} B. Halperin and M. Lax, Phys. Rev. {\bf 148}, 722
(1966).

\bibitem{ZL}  J.~Zittartz and J.S.~Langer, Phys. Rev. {\bf 148},
34 (1966).

\bibitem{Coleman} See, {\em e.g.}, S. Coleman, {\em ``Aspects of
symmetry''}, (Cambridge University Press, Cambridge, 1988).

\bibitem{Keldysh}L. V. Keldysh, Zh. Eksp. Theor. Fiz. {\bf 45},
364 (1963) [Sov. Phys. JETP {\bf 18}, 253 (1964)].

\bibitem{XT} B. C. Xu and S. E. Trulinger, Phys. Rev. Lett. {\bf
57}, 3113 (1986).

\bibitem{Kim} K. Kim, R. H. McKenzie, and J. H. Wilkins, Phys.
Rev. Lett. {\bf 71}, 4015 (1993).

\bibitem{H&M} R. Hayn and J. Mertsching, Phys. Rev. B {\bf 54},
R5199 (1996).

\bibitem{S&M} B. Starke and J. Mertsching, Synth. Met. {\bf 76},
217 (1996).

\bibitem{LSP} See, {\em e.g.}, I. M. Lifshits, S. A. Gredeskul
and L. A. Pastur, {\em ``Introduction to the theory of disordered
systems''} (Wiley Interscience, New York, 1988).

\bibitem{BK} S. A. Brazovski\u{i} and N. Kirova, Sov. Phys. JETP
Lett. {\bf 33}, 4 (1981).

\bibitem{CB} D. K. Campbell and A. R. Bishop, Phys. Rev. B {\bf
24}, 4859 (1981).

\bibitem{SSH} A. J. Heeger, S. Kivelson, J. R. Schrieffer, and
W.~P.~Su, Rev. Mod. Phys. {\bf 60}, 781 (1988).

\bibitem{TLM} H. Takayama, Y. R. Lin-Liu, and K. Maki, Phys. Rev.
{\bf 21}, 2388 (1980).

\bibitem{Thouless}, D. J. Thouless, J. Phys. C: Solid State
Phys., {\bf 5}, 77 (1972).

\bibitem{Dyson} F. J. Dyson, Phys. Rev. {\bf 92}, 1331 (1953).

\bibitem{ER} T. P. Eggarter and R. Riedinger, Phys. Rev. B {\bf
18}, 569 (1978) and references therein.



\end{references}
\end{document}